\documentstyle[pre,aps,epsf,preprint,tighten]{revtex}

\begin{document}

\title{
Correlations of chaotic eigenfunctions:
a semiclassical analysis
}
\author{
Baowen Li$^{1}$ \footnote{Email:\quad phylibw@nus.edu.sg} and Daniel C Rouben$^{2}$\footnote{Email:\quad rouben@aece.ca}
} 
\address{
$^{1}$ Department of Physics, National University of Singapore, 
117542 Singapore \\
$^{2}$Atomic Energy of Canada Limited, Mississauga, Ontario. CANADA L5K 1B2\\
}
\date{Published 31 August 2001, J. Phys. A {\bf 34}, 7381-7391(2001)} 
\maketitle

\date{\today}

\begin{abstract}

We derive a semiclassical expression for an energy smoothed autocorrelation
function defined on a group of eigenstates of the Schr\"odinger equation.
The system we considered is an energy-conserved Hamiltonian system
possessing time-invariant symmetry. The energy smoothed autocorrelation
function is expressed as a sum of three terms. The first one is analogous to Berry's
conjecture, which is a Bessel function of the zeroth order. The second and the third terms are  trace formulae made from special trajectories. The second term is found to be direction dependent in the case of spacing averaging, which agrees qualitatively with previous numerical observations in high-lying eigenstates of a chaotic billiard.
\end{abstract}


\maketitle

\section{Introduction}

With the intention of finding
a fingerprint of classical chaos in quantum systems, 
the study of eigenstates has attract increasing attentions both theoretically \cite{Shnirelman,Berry77,Voros79,MK88,Berry83,Heller84,Bogomolny88,Berry89,Prosen,Steiner93,Li94,Li95,Li96,Srednicki96,Li98,Steiner,Eckhardt,Srednicki98,Kaplan99,Li00}
and experimentally\cite{Sridhar00,Stockman} over the last two decades.
Among many others, the statistics of wavefunction amplitude $|\psi({\bf r})|^2$ and two-point correlation function $\langle\psi({\bf r}')\psi({\bf r}'')\rangle$ have been widely studied. Recently, the study has been extended to the statistics of the Ricaati variable of wave function, and a new universality has been found in chaotic and disordered systems\cite{Li00}.  The amplitude distribution of an eigenfunction in a classically chaotic quantum billiard system is conjectured to be Gaussian in the semiclassical limit by Berry\cite{Berry77}. The numerical studies in very high-lying eigenstates support this conjecture\cite{Steiner93,Li94,Li98}. 
In fact, in the far-semiclassical limit, even in scarred eigenstates, this conjecture remains true\cite{Li98}.

The correlation function of the wavefunction reveals some local fluctuation properties of the eigenstates. Following Berry's conjecture\cite{Berry77} that the wavefunction for a chaotic billiard is a superposition of infinite plane waves with random coefficients, the two-point correlation function is found to be proportional to $J_0(kd)$ ($J_0$ is the Bessel function of the zeroth order, $k$ the wave vector and $d$ the distance between the two points). The numerical calculations in high-lying chaotic eigenstates reveal some discrepancy with this conclusion both in a quantum system whose classical counterpart is the geodesic motion on a compact surface of constant negative curvature (as studied by Aurich and Steiner\cite{Steiner93}) and in the Lima\c con billiard (as studied by Li and Robnik\cite{Li94}). Later on, Srednicki and Stiernelof\cite{Srednicki96} investigated this fluctuation from a statistical point of view.

In a detailed study of the autocorrelation function of an individual eigenstate, Li and Robnik\cite{Li94} found that the two-point correlation function is direction dependent. Only after averaging over many directions, do the numerical calculations agree with Berry's prediction. To date, a satisfactory explanation for this discrepancy has been lacking. This motivated us to study analytically the two-point correlation function of eigenstates in a chaotic quantum billiard in terms of periodic orbit theory.
 
The billiard system we are considering is a quantum system of one particle moving in a closed domain ${\cal B}$ with area ${\cal A}$ and boundary 
$\partial {\cal B}$. The potential vanishes inside 
the billiard and has infinite value otherwise, 
$V({\bf r})=0$ for all ${\bf r}\in {\cal B}$ and
$V({\bf r})=\infty$ for all ${\bf r}\in \partial {\cal B}$.
The quantum eigenstate of such a billiard satisfies the Schr\"odinger 
equation
(Helmholtz equation),
\begin{equation}
\left(\frac{\hbar^2}{2m}\nabla+E\right )\psi({\bf r})=0,
\label{Schr}
\end{equation}
with Dirichlet boundary condition, $\psi({\bf r})=0$ for ${\bf r} \in \partial {\cal B}$, and is normalized, i.e.,
$\int_{\cal A}d{\bf r} |\psi ({\bf r})|^2 =1$.

In order to define a two-point correlation function we first 
consider a point ${\bf r}$ in some subset of ${\cal B}$, 
say ${\cal C}$, with area given by ${\cal A_C}$. We will
impose two conditions on ${\cal C}$. First, all points in
${\cal C}$ are said to be far from the boundary, 
$\partial {\cal B}$. Second, we fix ${\cal A_C}$ such that 
${\cal A_C}<<{\cal A}$. Third, we consider a real vector, 
${\bf d}$, such that for all ${\bf r}\in {\cal C}$, 
${\bf r}\pm {\bf d}/2$ lies within the classically allowed 
region but not necessarily in ${\cal C}$. We then introduce a function given by
\begin{equation}
C({\bf r},{\bf d};\Delta E)=\frac{
\sum_n\psi_n({\bf r}+ {\bf d}/2)\psi_n^*({\bf r} -{\bf d}/2)}{
\sum_n|\psi_n(\bf r)|^2}. 
\label{defa}
\end{equation}

We also consider the position-smoothed object given by
\begin{equation}
C({\cal C},{\bf d};\Delta E)=
\frac{\int_{\cal C} d{\bf r}\
\sum_n\psi_n({\bf r+ d}/2)\psi_n^*({\bf r - d}/2)} 
{\int_{\cal C} d{\bf r}\sum_n|\psi_n(\bf r)|^2}. 
\label{defb}
\end{equation}

Both definitions correspond to a two-point correlation function (the 
two points are (${\bf r}=\pm {\bf d}/2)$)) of a group of 
eigenstates. In the first case, it is a function of a point in 
${\cal C}$, ${\bf r}$, an energy window, $\Delta E$, and a 
real vector, ${\bf d}$; in the second case we perform a position 
smoothing in ${\bf r}$ over the region ${\cal C}$. Let $N$ be the 
number of eigenstates whose eigenenergies lie in $\Delta E$; all 
$N$ eigenstates must appear in the sum in equations (\ref{defa}) and (\ref{defb}). We also define an energy $E_0$ such that
$$
E_0-\frac{1}{2}\Delta E\le E_n \le E_0+\frac{1}{2}\Delta E
$$
for all $n$ in the set of eigenstates.

Thus, from the outset we consider an energy-averaged 
(energy-smoothed) object as in \cite{Bogomolny88} but may, 
in the end, consider the limit of small $\Delta E$ such that 
$N=1$. We will also mention that the correlation function with 
energy smoothing has also been widely used in the study of 
disordered systems (for a review please see Ref.\cite{Mirlin} 
and the references therein).

Historically the autocorrelation function in quantum chaos has 
revolved around the work of Berry. Berry's original definition 
\cite{Berry77} is similar to equation (\ref{defb}) but considers 
strictly one eigenstate and imposes a technical but important 
restriction on the area, ${\cal A_C}$. Berry considers the 
function given by
\begin{equation}
C({\cal C},{\bf d}; E= E_n)=\frac{\int_{\cal C} d{\bf r}\
\psi_n({\bf r+ d}/2)\psi_n^*({\bf r -d}/2)}
{\int_{\cal C} d{\bf r}|\psi_n({\bf r})|^2}
\label{Berrydef}
\end{equation}
with the area, ${\cal A_C}$ is mentioned in the following. 

From the outset, the quantity in equation (\ref{Berrydef}) is chosen to 
be a function of one eigenstate, $E_n$, rather than of an energy 
window containing one or more eigenstates. As for ${\cal C}$, it 
is chosen such that its volume tends to zero in the semiclassical 
limit\cite{Berry77}, i.e.
$\mbox{lim}_{\hbar\to 0}{\cal A_C}=0\;\;\; \mbox{but}\;\;\;
\mbox{lim}_{\hbar\to 0}\hbar/\sqrt{\cal A_C}=0.
$
As pointed out by Berry \cite{Berry83}, the vanishing of ${\cal A_C}$
ensures the smoothed functions to be `semiclassically sharp', whereas
$\sqrt{\cal A_C}$ vanishes more slowly than $\hbar$ so that the
oscillatory detail on the scale of the de Broglie wavelength is smoothed
away. This crucial point also allows one to reduce equation (\ref{Berrydef}) 
to a Bessel function, $J_0(kd)$, where $k=\sqrt{2mE}/\hbar$ is the 
de Broglie wave number. Li and Robnik's numerical calculations are based on
equation (\ref{Berrydef}) and the condition given above.

The goal of this paper is to shed light on the above-mentioned direction-dependence 
problem of Berry's autocorrelation function by (4) working with 
the closely related definitions, equations (\ref{defa}) and (\ref{defb}),
in the framework of periodic orbit theory. The 
introduction of an energy window, $\Delta E$, will permit us to do what we 
want in a clear manner. Moreover, by using standard semiclassical methods, 
we will be able to calculate a first term given by $J_0(kd)$ and a second 
term which carries all directional dependencies of the vector ${\bf d}$.

The paper is organized as follows. In section 2 we give a general semiclassical formula 
for two-point correlation function. In section 3, the effect of energy averaging is discussed. The effect of position and direction averaging is discussed in sections. 4 and 5, respectively. The paper is concluded by a summary and discussion in section 6.

\section{Semiclassical expression of two-point correlation function}

To develop equations (\ref{defa}) or (\ref{defb}), let us begin 
by introducing the Green function in the energy representation 
corresponding to equation (\ref{Schr}). It satisfies the following 
equation,
\begin{equation}
\left (E+\frac{\hbar^2}{2m}\nabla_{\bf r''}\right )
G({\bf r''},{\bf r'};E)=\delta({\bf r''}-{\bf r'}).
\end{equation}
Using the series expansion for $G({\bf r''},{\bf r'};E)$ in terms 
of the eigenstates of the Hamiltonian:
\begin{equation}
G({\bf r''},{\bf r'};E)=
\sum_n\frac{\psi^*_n({\bf r''})\,\psi_n({\bf r'})}{E-E_n+i\epsilon}
\label{GrEx}
\end{equation}
where $\epsilon>0$, it is easy to recast equation (\ref{defa}) in terms 
of the Green function,
\begin{equation}
\sum_n\psi_n({\bf r'})\psi^*_n({\bf r''})\delta(E-E_n) = 
\frac{1}{2\pi i} \left[G^*({\bf r'}, {\bf r''}, E) -G({\bf r''}, {\bf r'},
E)\right]
\end{equation}
Take the integral over energy integral $[E_0-\Delta E/2, E_0+\Delta E/2]$ on 
both sides we have:
\begin{equation}
\sum_n\psi_n({\bf r'})\psi^*_n({\bf r''}) = 
\frac{1}{2\pi i} \left[\langle 
G^*({\bf r'}, {\bf r''}, E) -G({\bf r''}, {\bf r'}, E) 
\rangle\right]
\end{equation}
where $\langle f(E)\rangle$ represents an energy smoothing or an integral on
the energy window, $\Delta E$, containing the $N$ energy levels,
$$
\langle f(E)\rangle=
\frac{1}{\Delta E}
\int_{E_0-\Delta E/2}^{E_0+\Delta E/2}\,dE'\,f(E').
$$

Thus we have:

\begin{equation}
\sum_n|\psi_n({\bf r})|^2 = 
\frac{1}{2\pi i} \left[\langle G^*({\bf r}, {\bf r}, E) -
G({\bf r}, {\bf r}, E)\rangle\right] = -\frac{1}{\pi}\Im m\langle
G({\bf r}, {\bf r}; E)\rangle,
\end{equation}

\begin{equation}
C({\bf r}, {\bf d};\Delta E) = \frac{1}{2i} \frac{ 
\langle G({\bf r} + {\bf d}/2, {\bf r} - {\bf d}/2;E)\rangle -
\langle G^*({\bf r} - {\bf d}/2, {\bf r} + {\bf d}/2;E)\rangle }
{\Im m\langle G({\bf r}, {\bf r}; E)\rangle}
\label{cora}
\end{equation}

and the position smoothed object,
\begin{equation}
C({\cal C},{\bf d};\Delta E) = \frac{1}{2i} \frac{\int_{\cal C} d{\bf r} 
\langle G({\bf r} + {\bf d}/2, {\bf r} - {\bf d}/2;E)\rangle -
\langle G^*({\bf r} - {\bf d}/2, {\bf r} + {\bf d}/2;E)\rangle }
{\int_{\cal C} d{\bf r} \Im m\langle G({\bf r}, {\bf r}; E)\rangle}
\label{corb}
\end{equation}

In the semiclassical limit, the Green function is decomposed
into a sum over two independent contributions:
\begin{equation}
G({\bf r''},{\bf r'};E)=G_{DI}({\bf r''},{\bf r'};E)
+G_{IN}({\bf r''},{\bf r'};E)+{\cal O}
\left (\frac{1}{\hbar}\right ).
\label{decomp}
\end{equation}

The first term $G_{DI}$ on the right represents a contribution from short-range
trajectories passing directly from ${\bf r'}$ to ${\bf r''}$. The 
second term $D_{IN}$ represents a long-range contribution (i.e. they make one
or more bounces from the billiard wall) from classical 
trajectories beginning and finishing at these points. 

Now we shall calculate the numerator and denominator in equations (\ref{cora}) 
and (\ref{corb}) separately. The denominator,

\begin{equation}
\Im m\langle G({\bf r}, {\bf r}; E)\rangle
=\Im m\langle G_{DI} ({\bf r},{\bf r}; E)\rangle 
+ \Im m\langle G_{IN} ({\bf r},{\bf r}; E)\rangle.
\label{Denom}
\end{equation}

Given,
$$
\Im m G_{DI} ({\bf r}, {\bf r}; E) = -\pi \int\delta[E-H({\bf p,r})] \frac{d 
{\bf p}}{(2\pi\hbar)^2}
$$
and that for a free particle moving inside the quantum billiard, 
$H({\bf p,r})={\bf p}^2/2m $, we have

\begin{equation}
\Im m \langle G_{DI}({\bf r}, {\bf r}; E)\rangle 
= -\frac{m}{2\hbar^2}
\label{resultat} 
\end{equation}

The contribution from long-range trajectories (indirect trajectories)
is given by the conventional sum over classical trajectories 
starting and ending at ${\bf r}$ 

\begin{equation}
G_{IN}({\bf r},{\bf r};E) = \frac{m}{i\hbar(2\pi i\hbar)^{1/2}}\sum_{cl}
|\Delta|^{1/2}
\exp{\left(\frac{i}{\hbar}S({\bf r},{\bf r};E)-
i\gamma \frac{\pi}{2}\right)},
\label{GrL}
\end{equation}

$$\Delta={\bf det}
\left (\frac{1}{|{\bf v''}||{\bf v'}|}
\frac{\partial^2S({\bf r''},{\bf r'};E)}{\partial t''\partial t'}
\right )_{t'=t''=0;{\bf r'} ={\bf r''}={\bf r}}.
$$
Here $\gamma$ is the count along the trajectory where the determinant 
($\Delta_d$) of second derivatives of the action in the pre-factor 
diverges. $t'(t'')$ measures a distance perpendicular to the trajectory 
at the initial (final) point and $v'(v'')$ is the velocity of the 
trajectory at the initial (final) point. The definition of the 
coordinates $t$ will become evident shortly. 

Equations (\ref{cora}) and (\ref{corb}) give the general formulae for two-point correlation function. In the subsequent sections we will evaluate these two formulae by different smoothing.

\section{The effect of energy averaging}

For the rest of the discussion we will write a subscript,
$d$, for all data referring to the trajectory, 
${\bf r}\pm {\bf d}/2\to {\bf r}\mp {\bf d}/2$, and $0$ for
all data referring to the closed trajectory, ${\bf r}\to {\bf r}$.
Thus, $S_0\equiv S_{cl}({\bf r}, {\bf r}; E)$ and
$S_d\equiv S_{cl}({\bf r}\pm {\bf d}/2,{\bf r}\mp {\bf d}/2;E)$. 
We have

\begin{equation}
\Im m \langle G_{IN}({\bf r}, {\bf r}; E)\rangle 
= - \frac{m}{\hbar\sqrt{2\pi\hbar}} \sum_{cl} |\Delta_0|^{1/2} 
\cos\left(\frac{S_0}{\hbar} -\frac{\pi}{4}(2\gamma+1)\right)
\end{equation}

Using equation (\ref{resultat}), the denominator (\ref{Denom}) becomes
\begin{equation}
\Im m\langle G({\bf r},{\bf r}; E)\rangle =
-\frac{m}{2\hbar^2} 
\left[ 
1 + \sqrt{\hbar}\frac{2}{\sqrt{2\pi}} \sum_{cl} |\Delta_0|^{1/2} 
\langle\cos\left(\frac{S_0}{\hbar} 
-\frac{\pi}{4}(2\gamma+1)\right)\rangle
\right] 
\label{Denom2}
\end{equation}

By the same procedure we can also calculate the numerator, we denote it 
as $I_N$,
\begin{eqnarray}
I_N & = & 
\langle 
G({\bf r} + {\bf d}/2, {\bf r} - {\bf d}/2; E) - 
G^*({\bf r} - {\bf d}/2, {\bf r} + {\bf d}/2; E)
\rangle\nonumber\\
& = & I_N^{DI} +  I_N^{IN}
\label{Nom1}
\end{eqnarray}

where
\begin{eqnarray}
I_N^{DI} & = & \langle G_{DI}({\bf r} + {\bf d}/2, {\bf r} - {\bf d}/2; E) 
- G_{DI}^*({\bf r} - {\bf d}/2, {\bf r} + {\bf d}/2; E)\rangle\nonumber\\
I_N^{IN} & = & \langle 
G_{IN}({\bf r} + {\bf d}/2, {\bf r} - {\bf d}/2; E) - 
G_{IN}^*({\bf r} - {\bf d}/2, {\bf r} + {\bf d}/2; E)\rangle
\end{eqnarray}

For a quantum billiard the short-range Green function is given by the 
usual formula,
\begin{equation}
G_{DI}({\bf r''},{\bf r'}; E)=\frac{m}{2i\hbar^2}H_{0}(k|{\bf r''}-{\bf r'}|)
\end{equation}
where $H_{0}(x)$ is the zeroth order Hankel
function of the first kind and $p=|{\bf p}|=\sqrt{2mE}$, $k=p/\hbar$.
Posing ${\bf r''}={\bf r+d/2}$ and ${\bf r'}={\bf r-d/2}$ we get
\begin{equation}
G_{DI}({\bf r\pm d/2},{\bf r\mp d/2},E) = \frac{m}{2i\hbar^2}H_{0}(kd).
\label{Hankel}
\end{equation}

Note that this Green function depends only on the position coordinates 
via $d=|{\bf r''}-{\bf r'}|$ and satisfies the natural boundary 
condition. The first term in equation (\ref{Nom1}) then becomes,

\begin{equation}
I_N^{DI} = -i\frac{m}{\hbar^2}\langle J_{0}(kd)\rangle
\end{equation}
and the second term,
\begin{equation}
I_N^{IN} = 
\frac{2m}{i\hbar\sqrt{2\pi\hbar}} \sum_{cl} |\Delta_d|^{1/2} 
\langle\cos\left(\frac{S_d}{\hbar} -\frac{\pi}{4}(2\gamma+1)\right)
\rangle
\end{equation}

Thus,
\begin{equation}
I_N = \frac{m}{i\hbar^2} \left[\langle J_0(kd)\rangle + 
\sqrt{\hbar}\frac{2}{\sqrt{2\pi}} \sum_{cl} |\Delta_d|^{1/2} 
\langle\cos\left(\frac{S_d}{\hbar} -\frac{\pi}{4}(2\gamma+1)\right)
\rangle
\right]
\label{Num2}
\end{equation}

We should point out that to obtain the above result, 
we have used the system's time-reversal symmetry whereby the 
trajectory, ${\bf r+d/2}\to {\bf r-d/2}$, has the same classical 
properties entering equation (\ref{Nom1}) as the trajectory 
${\bf r-d/2}\to {\bf r+d/2}$.

Substituting the denominator, equation (\ref{Denom2}), and numerator,
equation (\ref{Num2}), together into 
 equation (\ref{decomp}) we have finally an expression for 
equation (\ref{defa}) up to the leading term of $\sqrt{\hbar}$,

\begin{eqnarray}
C({\bf r},{\bf d};\Delta E) = &&\langle J_{0}(kd)\rangle 
+
\sqrt{\hbar}\sqrt{\frac{2}{\pi}}\sum_{cl} |\Delta_d|^{1/2}
\langle 
\cos\left(\frac{S_{d}}{\hbar} -\frac{\pi}{4} (2\gamma +1)\right)
\rangle\nonumber\\
& -& \sqrt{\hbar} 
\sqrt{\frac{2}{\pi}}\sum_{cl} 
|\Delta_0|^{1/2}
\langle\cos\left(\frac{S_0}{\hbar} 
-\frac{\pi}{4}(2\gamma +1)\right)
\rangle 
\label{finale1}
\end{eqnarray}

The sum in the second term in the above equation is over all 
trajectories connecting ${\bf r+d/2}$ and ${\bf r-d/2}$. 
In the third term it is over all closed trajectories starting
and finishing at ${\bf r}$. The energy smoothing has the 
subsequent affect of giving preference to trajectories with a 
time of motion satisfying, $\tau<\hbar/\Delta E$. Thus as we 
reduce the value of $\Delta E$ we are forced to add an increasing 
number of trajectories in equation (\ref{finale1}) to maintain 
numerical consistency and accuracy. Conversely, by choosing 
an average over many states we are free to choose less trajectories 
in the sum above, an evident advantage from the numerical 
point of view.

\section{The effect of  position averaging}

Now we consider the case where numerator and denominator in
equation (\ref{defa}) are to be integrated over some small region 
${\cal C}$ with area, ${\cal A_C}$. This amounts to integrating 
equation (\ref{Denom2}) and equation (\ref{Nom1}) over ${\cal C}$. In order to
proceed we will first consider the possibility that $A_{\cal C}$ is 
either fixed (in $\hbar$) or goes to zero so slowly in the 
semiclassical limit that the long-range Green function has many 
rapid oscillations on ${\cal C}$.

With the above-mentioned simplification, the long-range 
Green function can be integrated via the method of stationary 
phase. The integral of interest has the general form,
$$
\int\;dx\int\;dy\;\;
\sum_{cl}e^{iS_{cl}({\bf r\pm d}/2,{\bf r\mp d}/2;E)/\hbar}.
$$
The stationary phase points in ${\cal C}$, labeled by 
${\bf\bar r}$, are solutions of the equation,
\begin{equation}
\left (\frac{\partial S}{\partial {\bf r''}}\right )_{({\bf\bar r\pm d}/2,
{\bf\bar r\mp d}/2)}+
\left (\frac{\partial S}{\partial {\bf r' }}\right )_{({\bf\bar r\pm d}/2,
{\bf\bar r\mp d}/2)}=0.
\label{condition}
\end{equation}
The solutions of equation (\ref{condition}) correspond to points ${\bf\bar r}$ 
in ${\cal C}$ for which there exists a trajectory leaving a point
${\bf r'}={\bf\bar r\pm d}/2$ and arriving at a point
${\bf r'}={\bf\bar r\mp d}/2$
with equal initial and final momenta in $x$ and $y$. This translates into a
search for points ${\bf\bar r}$ in ${\cal C}$ affiliated with a trajectory which 
has the following two properties:
\begin{itemize}
\item (a) its path includes two parallel chords
\item (b) on one chord there exists a point, ${\bf\bar r}\pm {\bf d}/2$,
and on the other a point, ${\bf\bar r}\mp {\bf d}/2$.
\label{crit}
\end{itemize}
The first condition assures that the initial and final momenta are equal.
The second condition assures that a vector, ${\bf d}$, can be oriented
between the two chords with midpoint given by ${\bf\bar r}$. Geometrically,
we search for a trajectory whose path includes two parallel chords that lie
on opposite sides (the perpendicular distance is given by ${\bf d}_{\perp}/2$)
of a third chord also running parallel to the two outer chords and lying in 
${\cal C}$ (please see Fig.1). It is not important that the two chords lie 
in $\cal C$, however, by definition ${\bf\bar r}$ must lie on the midpoint
chord which lies in this region. It is also clear that the solutions
of equation (\ref{condition}) are not isolated; in fact they lie on a line, the
midpoint chord. We speak of a line of stationary phase points, $S$, and to 
treat the integration in $x$ and $y$ we introduce the usual system of 
coordinates to describe motion local to a trajectory. Moreover, due to 
time-invariance a solution corresponding to the path 
${\bf\bar r+d/2}\to {\bf\bar r-d/2}$ will also be a solution corresponding to
the path ${\bf\bar r-d/2}\to {\bf\bar r+d/2}$.

\begin{figure} 
\epsfxsize=11cm
\epsfbox{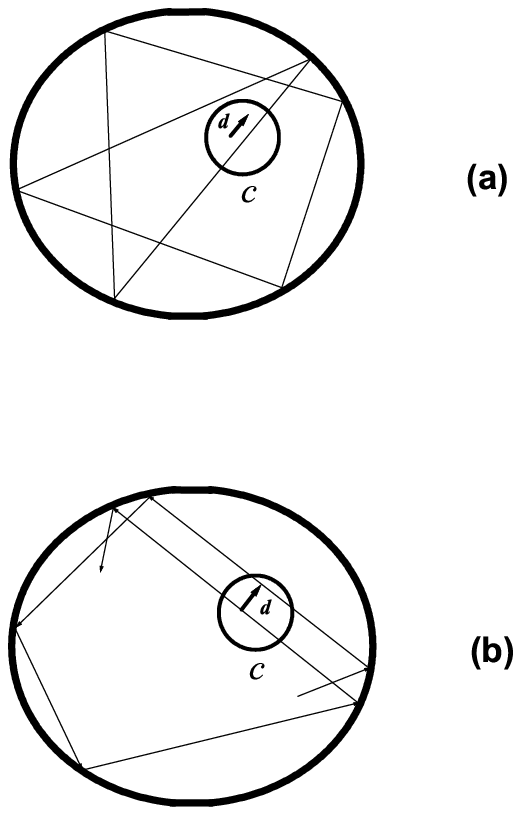}
\vspace{-1.5cm}
\caption{Schematic illustration of two different kinds of trajectories contributing to the second term on the right-hand side of equation (38). (a) Periodic orbit which has a chord in ${\cal C}$ parallel to ${\bf d}={\bf r}''-{\bf r}'$. (b) Non periodic orbit which has two segments parallel to each other and passing through the two ends of ${\bf d}$. It is obvious that the contribution depends on the direction of ${\bf d}$.}
\label{fig1}
\end{figure}

There exist two types of trajectories that satisfy the required conditions
(see Fig.~(\ref{fig1})). Trivially any periodic orbit which has a chord in
${\cal C}$ parallel to ${\bf d}$ will satisfy this condition and its
stationary phase points then lie on that chord (Fig. 1a). A second possibility comes
from non-periodic trajectories (Fig. 1b). In both cases $v''=v'=\dot s$. Having
found the stationary phase points, one can expand the action to second order in 
both $x$ and $y$. On the other hand, for the purposes of integrating over 
${\cal C}$ it is most convenient to introduce the usual Cartesian set 
of local coordinates, describing motion parallel and perpendicular to $S$. 
Let us denote this new set as $(s,t)$ with 
$|\partial{(x,y)}/\partial{(s,t)}|=1$.
The variable $t$ measures a displacement perpendicular to $S$ and $s$ 
measures a displacement along $S$. For case (1) trajectories the point 
${\bf\bar r}$ lies on the trajectory itself but in the second case it 
lies between two chords that are parallel to one another (and to $S$) 
and are separated by ${\bf d}_{\perp}/2$. 

For both cases (1) and (2), one now proceeds to integrate over $s$ and $t$
by expanding the action to second order in $t$ (in both initial and final point), 
for fixed $s$, and integrating first in $t$. Expansion of the action to second 
order in $t$ gives,
\begin{eqnarray}
S((s+d_{\parallel}/2,&t&+d_{\perp}/2),(s-d_{\parallel}/2,t-d_{\perp}/2);E)
\nonumber\\
=&S&((s+d_{\parallel}/2,d_{\perp}/2),(s-d_{\parallel}/2,-d_{\perp}/2);E)
\nonumber\\
&+&\frac{1}{2}
\left (
\frac{\partial^2 S}{\partial t^{'2}}+
\frac{\partial^2 S}{\partial t^{''2}}+2
\frac{\partial^2 S}{\partial t'\partial t''}
\right )t^2 + {\it O}(t^3).
\label{expansion}
\end{eqnarray}
The first term refers to the action, $S_d$, for going from the
point $(s+d_{\parallel}/2,d_{\perp}/2)$ to the point
$(s-d_{\parallel}/2,-d_{\perp}/2)$. It is clearly invariant in $s$, the
coordinate along $S$. Fixing $s$ at one value on $S$, we can define 
the linear Poincar\'e map, $d{\it P}((t',p')\to (t'',p''))$. It is 
a matrix with four elements given by:
\begin{eqnarray}
\delta t''&=&m_{11}\delta t'+m_{12}\delta p'\nonumber\\
\delta p''&=&m_{21}\delta t'+m_{22}\delta p'.
\label{monodromie}
\end{eqnarray}

We also have the well known relations \cite{GutzB}
\begin{equation}
\frac{\partial^2 S}{\partial t^{'2}}=\frac{m_{11}}{m_{12}},\;\;\;
\frac{\partial^2 S}{\partial t^{''2}}=\frac{m_{22}}{m_{12}},\;\;\;
\frac{\partial^2 S}{\partial t'\partial t''}=-\frac{1}{m_{12}}
\label{relations}
\end{equation}
which we use to write the second variation as $(m_{11}+m_{22}-2)/(2m_{12})$.
Again in equation (\ref{monodromie}) we are considering variations in $t$ at the
initial and final points.

The indirect part of the Green function is given by \cite{GutzB},

\begin{equation}
G_{IN}({\bf r}'',{\bf r}';E) = \frac{m}{i\hbar(2\pi i\hbar)^{1/2}}\sum_{cl}
\frac{|D(s)|^{1/2}}{|v|}
\exp{\left(\frac{i}{\hbar}\left(S_d+\frac{W(s)}{2}t^2\right) 
-i\gamma \frac{\pi}{2}\right)},
\label{Nonsm}
\end{equation}

where
\begin{equation}
D(s) = \frac{1}{m^d_{12}}, \qquad\, W(s) = 
\frac{m^d_{11} + m^d_{22}- 2}{m^d_{12}}.
\label{monodrom2}
\end{equation}
Again the symbol $d$ refers to classical motion beginning at
${\bf\bar r}\pm {\bf d}/2$ and finishing at 
${\bf\bar r}\mp {\bf d}/2$ whereas the symbol $0$ refers to the 
special case, ${\bf d}=0$.

The integration over $t$ gives rise to

\begin{equation}
\int_{-\infty}^{+\infty} dt \exp{\left(\frac{iW(s)}{2\hbar}t^2 \right)}
=\sqrt{\frac{2\pi\hbar}{|m^d_{11} + m^d_{22} -2|}} 
\exp{\left(i\frac{\pi}{4} sign\left(m^d_{11}+ m^d_{22}-2\right)\right)}
\end{equation}

Here the sign function is given by;

\begin{equation}
sign(x) = \left\{\begin{array}{ll} 
1 & x>0 \\ 
-1 & x<0
\end{array}
\right.
\label{eqtype2}
\end{equation}

Due to the invariance of action and the term $m_{11}+m_{22}$ the 
subsequent integration in $s$ is particularly simple; from it we 
get the length of $S$ in ${\cal C}$ divided by the speed along it, 
$|v|$, in other words a time which we denote $T^d_{cl}$. This gives 

\begin{equation}
\int_{\cal A_C} d{\bf r} \langle G_{IN}({\bf r} \pm {\bf d}/2, {\bf 
r}\mp{\bf d}/2,E)\rangle =-i\sqrt{\hbar}m
\sum_{cl} \frac{T_{cl}}{\sqrt{|m^d_{11}+m^d_{22}-2|}}
\langle \exp{\left(\frac{i}{\hbar}S_d - \frac{i}{4}\pi\sigma_d 
\right)}\rangle
\end{equation}

where

$$
\sigma_d = 2\gamma_d + 1 -\sigma(m^d_{11} + m^d_{22} -2)
$$

Therefore,

\begin{eqnarray}
\int_{\cal A_C} & & d{\bf r} \langle G_{IN}({\bf r} + {\bf d}/2, {\bf 
r} - {\bf d}/2,E) - G_{IN}^*({\bf r} - {\bf d}/2, {\bf 
r} + {\bf d}/2,E) \rangle\nonumber\\
& = & -i\sqrt{\hbar} m \sum_{cl} 
\frac{T^d_{cl}}{\sqrt{|m^d_{11}+m^d_{22} -2|}}\langle \cos 
\left(\frac{S_d}{\hbar} -\frac{\sigma_d}{4}\pi\right)\rangle
\end{eqnarray}

The numerator is thus,

\begin{equation}
I_N =\frac{m{\cal A_C}}{i\hbar^2} \left[ \langle J_0(kd) \rangle + 
\frac{2\hbar}{\cal A_C}\sum_{cl} 
\frac{T^d_{cl}}{\sqrt{|m^d_{11}+m^d_{22} -2|}}\langle \cos 
\left(\frac{S_d}{\hbar} -\frac{\sigma_d}{4}\pi\right)\rangle\right]
\end{equation}

Similarly, we can also obtain the expression for the denominator

\begin{equation}
\int_{\cal C} d{\bf r} \Im m\langle G({\bf r}, {\bf r}; E)\rangle =
-\frac{m{\cal A_C}}{2\hbar^2} \left[1 + \frac{\hbar}{\cal A_C} 
\sum_{cl} 
\frac{T^0_{cl}}{\sqrt{|m^0_{11}+m^0_{22} -2|}}\langle \cos 
\left(\frac{S_0}{\hbar} -\frac{\sigma_0}{4}\pi\right)\rangle\right]
\end{equation}

Finally the correlation function is given by,

\begin{eqnarray}
C({\cal C},{\bf d};\Delta E) = 
\langle J_0(kd)\rangle &+& \hbar\frac{2}{\cal A_C} 
\sum_{cl} 
\frac{T^d_{cl}}{\sqrt{|m^d_{11}+m^d_{22} -2|}}\langle \cos 
\left(\frac{S_d}{\hbar} -\frac{\sigma_d}{4}\pi\right)\rangle
\nonumber\\
& -& 
\hbar\frac{2}{\cal A_C}
\sum_{cl} 
\frac{T^0_{cl}}{\sqrt{|m^0_{11}+m^0_{22} -2|}}\langle \cos 
\left(\frac{S_0}{\hbar} -\frac{\sigma_0}{4}\pi\right)\rangle
\label{finale2}
\end{eqnarray}

The sum in the second term is over all trajectories satisfying 
the criteria (a) and (b). {\it It carries the directional dependency 
of ${\bf d}$ via the choice of trajectories}. The sum in the third 
term is over all periodic orbits starting and ending at the stationary
point in the averaging 
area of $\cal A_C$, this term does not depend on the direction.
All trajectories must be included, however, the subsequent energy 
smoothing will restrict this sum to trajectories whose time scale is 
$\tau\approx\hbar/\Delta E$. Again our final answer cannot depend 
on $\Delta E$ but this factor will control the contribution of each 
trajectory in the sum. Trajectories whose time-scales greatly exceed 
$\tau$ will be washed out by the energy smoothing. Any periodic orbit 
which has a chord parallel to ${\bf d}$ in ${\cal C}$ will contribute 
to this term. However, a periodic orbit which scars a particular state 
whose energy is outside the window, $\Delta E$, may not contribute.


\section{The effect of direction averaging}

In this section, we will consider the sum in
equation (\ref{finale2}) over different directions of the vector 
${\bf d}$. To do this one can picture a quantum billiard 
with vector ${\bf d}$ centered at $(x,y)=(0,0)$. Rotation of 
${\bf d}$ produces a circle of radius
$d/2$ centered at $(0,0)$. Each trajectory in equations (\ref{finale1}) and/or 
(\ref{finale2})
begins and finishes on this circle. The initial and final point
are separated by the diameter $d$. Most generally, the dynamics 
defines a map $P$ which is a function
of initial point ${\bf q}=(x,y)$, momentum ${\bf p}=(p_x,p_y)$
and time. The linear map ($DP$) is a 4*4 matrix.

To enumerate trajectories one constructs a symbolic dynamics 
\cite{Backer}. This construction allows one to assign to each
trajectory in equations (\ref{finale1}) and/or (\ref{finale2}) , a sequence of numbers 
(e.g. a code (i)) and a smooth function $S_i$, the action of 
that trajectory. Given a trajectory labeled by $i$ which 
starts at $\theta^*$ and finishes at $\theta^*+\pi$, the
idea now is to find a new trajectory which runs almost the 
same pattern but begins at 
$\theta=\theta^* + d\theta$ and finishes at 
$\theta '=\theta + \pi$. This matching can be performed 
by using the linear map, $DP$ (with matrix elements given by the
original trajectory), substituting
$\delta x=-d\sin\theta\delta\theta$, and
$\delta y= d\cos\theta\delta\theta$, and
solving for initial $(\delta p_x,\delta p_y)$. This method for 
matching can be continued until the linear map has a 
singular point. Clearly $S_i$, the action built for code i, is 
a function of $\theta$ for all $i$.  

Let us now denote terms (1),(2) and (3) in equations (\ref{finale1}) and/or  
(\ref{finale2}) by $f_{\circ}, f_1$ and $f_2$, respectively. We 
also define an average over $\theta$ given by 
$$
\tilde{f}(\theta_1,\theta_2)=\frac{1}{\theta_2-\theta_1}\int_{\theta_1}^{\theta_2}f 
d\theta. $$ 
Clearly $f_{\circ}$ and $f_2$ are independent of $\theta$ 
and can be trivially integrated; they remain of order $1$ and 
$\sqrt{\hbar}$ respectively. On the other hand, on averaging 
$f_1$ in the semiclassical limit, only points  $\theta^*$ where 
$\partial S_i/\partial\theta=0$ will give useful results.  After 
Gaussian integration, any term labeled by (i) in $\tilde{f}_1$, 
will  either be of order $\hbar$ if $\theta^*$ exists in the 
domain of integration  or zero otherwise. Thus in the semiclassical 
limit a sum over many $\theta$  would semiclassically clear away 
corrections of order $\sqrt{\hbar}$ which are $\theta$ dependent 
(e.g. the term $f_1$). Moreover, note that $\theta_1$ and  
$\theta_2$ need not satisfy $\theta_2-\theta_1=2\pi$.  

In the paper\cite{Li94}, numerical work on a single high-lying  
eigenfunction in the Lima\c con billiard was performed. It was 
shown there that while Berry's conjecture was not satisfied for 
a fixed $\theta$, a direction average over many $\theta$  
(with $0\le\theta <2\pi$) appeared to give a result which exactly 
corresponded to that of Berry's conjecture. This latter point 
would intuitively suggest that $\tilde{f}_1$ exactly cancels  
$\tilde{f}_2$ using $\theta_1=0$ and $\theta_2=2\pi$, 
$$ \frac{1}{2\pi}\int_0^{2\pi}f_1(\theta)d\theta=-\tilde{f}_2=-f_2, 
$$  leaving $f_{\circ}=<J_0(kd)>$ as the only surviving 
contribution. But according to our arguments, $\tilde{f}_1$ is of 
order $\hbar$ while  $\tilde{f}_2$ is of order $\sqrt{\hbar}$. 
Hence these two terms cannot  cancel each other. However, 
since $f_1$ is washed away by the direction  average, the 
differences from  $\tilde{f}_{\circ}=f_{\circ}=<J_{\circ}>$ will 
come from $f_2$ alone.


\section{Conclusions}

We have calculated a semiclassical expression for an (energy) averaged autocorrelation   
function over a group of eigenstates within the energy window $\Delta E$ up to the leading term of $\sqrt{\hbar}$. The final expression equations (\ref{finale1}) and (\ref{finale2}) are our main results for correlation functions equations (\ref{defa}) and (\ref{defb}), respectively.  
In both cases, the autocorrelation functions are expressed as a sum of three terms, the first one is the Bessel function of the zeroth oder (Berry's term), the second and the third ones are of special trajectory modulations. In the first case, i.e. definition equation (\ref{defa}), the second term (in equation (\ref{finale1})) includes the contribution from all trajectories connecting the two ends of the vector ${\bf d}={\bf r}'' -{\bf r}'$. The third term comes from the contribution of all {\it closed} trajectories starting and ending at ${\bf r}$. However, in the second case, i.e. have space averaging (equation (\ref{finale2}), things become quite different. The second term on the right-hand side of equation (\ref{finale2}) is the contribution from all trajectories satisfying the criteria (a) and (b). This term carries the directional dependency of ${\bf d}$. There are two kinds of trajectories satisfy these two criteria: (1) any periodic orbit which has a chord in the averaging domain ${\cal C}$ parallel to ${\bf d}$  (see figure 1(a)) makes contribution; (2) any open trajectory (see figure 1(b)) having two segments parallel to each other and passing through (not necessarily perpendicular to the vector ${\bf d}$) the two ends of ${\bf d}$ in the same direction contribute to this term. The averaging over all directions might wash away this term as observed numerically by Li and Robnik\cite{Li94}. The third term on the right-hand side of equation (\ref{finale2}) includes all periodic orbits starting and ending at the stationary point in the averaging domain ${\cal C}$.  In principle, the number of trajectories to be calculated is infinite, however, the energy smoothing introduced from beginning restricts this to trajectories and/or periodic orbits whose time scale is about $\tau \approx \hbar/\Delta E$. This gives an advantage in numerical calculation.


\begin{acknowledgments}
We would like to thank Eugen Bogomolny, Niall Whelan, Denis Ullmo and Wei-Mou Zheng
for helpful discussions, and the two anonymous referees for very stimulating comments and suggestions. BL was supported by the Academic Research Fund of National University of Singapore.
\end{acknowledgments}

\end{document}